


\magnification = 1200
\overfullrule=0pt

\font\titlerm = cmr10 scaled\magstep 4
\font\titlerms = cmr7 scaled\magstep 4
\font\titlermss = cmr5 scaled\magstep 4
\font\titlei = cmmi10 scaled\magstep 4
\font\titleis = cmmi7 scaled\magstep 4
\font\titleiss = cmmi5 scaled\magstep 4
\font\titlesy = cmsy10 scaled\magstep 4
\font\titlesys = cmsy7 scaled\magstep 4
\font\titlesyss = cmsy5 scaled\magstep 4
\font\titleit = cmti10 scaled\magstep 4

\def\titlefont{\def\rm{\fam0\titlerm}
\def\it{\fam\itfam\titleit}
\textfont0 = \titlerm
\scriptfont0 = \titlerms
\scriptscriptfont0 = \titlermss
\textfont1 = \titlei
\scriptfont1 = \titleis
\scriptscriptfont1 = \titleiss
\textfont2 = \titlesy
\scriptfont2 = \titlesys
\scriptscriptfont2 = \titlesyss
\textfont\itfam = \titleit
\rm}

\def\sectionfont{\def\rm{\fam0\tenrm}
\def\it{\fam\itfam\tenit}
\def\bf{\fam\bffam\tenbf}
\textfont0 = \tenrm
\scriptfont0 = \sevenrm
\scriptscriptfont0 = \fiverm
\textfont1 = \teni
\scriptfont1 = \seveni  \scriptscriptfont1=\fivei
\textfont2 = \tensy
\scriptfont2 = \sevensy
\scriptscriptfont2 = \fivesy
\textfont\itfam = \tenit
\textfont\bffam = \tenbf
\rm}

\font\teenyfont = cmr5

\global\baselineskip = 1.2\baselineskip
\global\parskip = 4pt plus 0.3pt
\global\nulldelimiterspace = 0pt

\predisplaypenalty 1000


\def\endignore{}
\def\ignore #1\endignore{}

\newcount\dflag
\dflag = 0


\def\monthname{\ifcase\month
\or Jan \or Feb \or Mar \or Apr \or May \or June%
\or July \or Aug \or Sept \or Oct \or Nov \or Dec
\fi}




\def\endid{}
\def\id#1\endid{\hfill #1}

\def\endtitle{}
\def\title#1\endtitle{\vskip.15in\titlefont
\global\baselineskip = 2\baselineskip
#1\vskip.3in
\baselineskip = 0.5\baselineskip\sectionfont}

\def\lblfoot{This work was supported by the Director, Office of Energy
Research, Office of High Energy and Nuclear Physics, Division of High
Energy Physics of the U.S. Department of Energy under Contract
DE-AC03-76SF00098.}

\def\endauthors{}
\def\authors#1\endauthors{
#1\if\dflag = 0
\footnote{}{\noindent\lblfoot}\fi}

\def\endabstract{}
\def\abstract#1\endabstract{\vskip .2in%
\centerline{\sectionfont\bf Abstract}%
\vskip .1in%
\noindent#1%
\ifnum\dflag = 0
\footline = {\hfil}\pageno = 0
\vfill\eject
\pageno = 1\footline{\centerline{\sectionfont\folio}}
\fi\ifnum\dflag = 2
\footline = {\hfil}\pageno = 0
\vfill\eject
\fi}


\newcount\nsection
\newcount\nsubsection

\def\section#1{\global\advance\nsection by 1
\global\nsubsection = 0
\bigskip\noindent
\centerline{\sectionfont\bf\number\nsection.\ #1}
\nobreak\medskip\sectionfont\nobreak}

\def\subsection#1{\global\advance\nsubsection by 1
\bigskip\noindent
\centerline{\sectionfont \it \number\nsection.\number\nsubsection.\ #1}
\nobreak\smallskip\rm\nobreak}

\def\appendix#1#2{\bigskip\noindent%
\sectionfont \bf Appendix #1.\ #2
\nobreak\medskip\rm\nobreak}


\newcount\nref
\global\nref = 1

\def\ref#1#2{\xdef #1{[\number\nref]}
#1
\ifnum\nref = 1\global\xdef\therefs{\noindent[\number\nref] #2\ }
\else
\global\xdef\oldrefs{\therefs}
\global\xdef\therefs{\oldrefs\vskip.1in\noindent[\number\nref] #2\ }%
\fi%
\global\advance\nref by 1
}

\def\listrefs{\vfill\eject\section{References}\therefs}


\newcount\nfig
\global\nfig = 1

\def\fg#1\efig{\vskip .5in\noindent Fig.\ \number\nfig:\ #1%
\global\advance\nfig by 1}


\newcount\cflag
\newcount\nequation
\global\nequation = 1
\def\eqlabel{(1)}

\def\nexteqno{\ifnum\cflag = 0
\global\advance\nequation by 1
\fi
\global\cflag = 0
\xdef\eqlabel{(\number\nequation)}}

\def\lasteqno{\global\advance\nequation by -1
\xdef\eqlabel{(\number\nequation)}}

\def\label#1{\xdef #1{(\number\nequation)}
\ifnum\dflag = 1
{\escapechar = -1
\xdef\draftname{\teenyfont\string#1}}
\fi}

\def\clabel#1#2{\xdef\eqlabel{(\number\nequation #2)}
\global\cflag = 1
\xdef #1{\eqlabel}
\ifnum\dflag = 1
{\escapechar = -1
\xdef\draftname{\string#1}}
\fi}

\def\cclabel#1#2{\xdef\eqlabel{#2)}
\global\cflag = 1
\xdef #1{\eqlabel}
\ifnum\dflag = 1
{\escapechar = -1
\xdef\draftname{\string#1}}
\fi}


\def\eeq{}

\def\eqnn #1\eeq{$$ #1 $$}

\def\eq #1\eeq{\xdef\draftname{\ }
$$ #1
\eqno{\eqlabel \rlap{\ \draftname}} $$
\nexteqno}



\def\eol{& \eqlabel \rlap{\ \draftname} \crcr
\nexteqno
\xdef\draftname{\ }}

\def\eeol{& \eqlabel \rlap{\ \draftname}
\nexteqno
\xdef\draftname{\ }}

\def\eolnn{\cr
\global\cflag = 0
\xdef\draftname{\ }}


\def\eqa #1\eeq{\xdef\draftname{\ }
$$ \eqalignno{ #1 } $$
\global\cflag = 0}


\def\ie{{\it i.e.\/}}
\def\eg{{\it e.g.\/}}

\def\etc{{\it etc}}

\def\myinstitution{
    \centerline{\it Theoretical Physics Group}
    \centerline{\it Lawrence Berkeley Laboratory}
    \centerline{\it 1 Cyclotron Road}
    \centerline{\it Berkeley, California 94720}
}


\def\jref#1#2#3#4{{\it #1} {\bf #2}, #3 (#4)}

\def\NPB#1#2#3{\jref{Nucl.\ Phys.}{B#1}{#2}{#3}}
\def\PA#1#2#3{\jref{Physica}{#1A}{#2}{#3}}
\def\PLB#1#2#3{\jref{Phys.\ Lett.}{#1B}{#2}{#3}}
\def\PR#1#2#3{\jref{Phys.\ Rep.}{#1}{#2}{#3}}
\def\PRD#1#2#3{\jref{Phys.\ Rev.}{D#1}{#2}{#3}}

\def\PRL#1#2#3{\jref{Phys.\ Rev.\ Lett.}{#1}{#2}{#3}}
\def\PRSL#1#2#3{\jref{Proc.\ R.\ Soc.\ London}{#1}{#2}{#3}}


\def\to{\mathop{\rightarrow}}
\def\too{\mathop{\longrightarrow}}


\def\frac#1#2{{{#1} \over {#2}}\,}  
\def\sfrac#1#2{{\textstyle\frac{#1}{#2}}}  


\def\Dsl{\hbox{/\kern-.6000em\rm D}} 



\def\twi{\widetilde}

\def\scr#1{{\cal #1}}
\def\op#1{{\widehat #1}}
\def\mybar#1{\kern 0.8pt\overline{\kern -0.8pt#1\kern -0.8pt}\kern 0.8pt}
\def\sla#1{\raise.15ex\hbox{$/$}\kern-.57em #1}
\def\Sla#1{\kern.15em\raise.15ex\hbox{$/$}\kern-.72em #1}

\def\roughly#1{\mathrel{\raise.3ex\hbox{$#1$\kern-.75em%
    \lower1ex\hbox{$\sim$}}}}
\def\lsim{\roughly<}


\def\tr{\mathop{\rm tr}}

\def\det{\mathop{\rm det}}


\def\bra#1{\langle #1 |}
\def\ket#1{| #1 \rangle}
\def\braket#1#2{\langle #1 | #2 \rangle}



\hyphenation{ba-ry-on ba-ry-ons ano-ma-ly ano-ma-lies}

\def\al{\alpha}
\def\del{\delta}
\def\Del{\Delta}
\def\gam{\gamma}
\def\Gam{\Gamma}
\def\ep{\epsilon}
\def\lam{\lambda}
\def\Lam{\Lambda}

\def\sig{\sigma}

\def\ChPT{\raise.45ex\hbox{$\chi$}PT}

\def\rhs{right-hand side}
\def\lhs{left-hand side}

\def\hc{{\rm h.c.}}


\def\MeV{{\rm \ MeV}}

\hyphenation{ba-ry-on ba-ry-ons}

\def\sf{$SU(2)_v \times SU(N_F)$}  

\def\bb{\bra{\scr B_0}}
\def\bk{\ket{\scr B_0}}

\def\bpb{\bra{\scr B'_0}}
\def\bpk{\ket{\scr B'_0}}

\def\rbra#1{(#1|}
\def\rket#1{|#1)}

\def\no#1{\mathopen{:}#1\mathclose{:}}
\def\op#1{\mathord{\{#1\}}}


\id
LBL-34778, hep-ph/9310369
\endid

\title
\centerline{Baryons from Quarks}
\centerline{in the $1 / N$ Expansion}
\endtitle

\authors
\centerline{Markus A. Luty\ \ {\it and}\ \ John March--Russell}
\footnote{}{\hskip-.27in\lblfoot}
\vskip .1in
\myinstitution
\endauthors

\abstract
We present a diagrammatic analysis of baryons in the $1/N$ expansion,
where $N$ is the number of QCD colors.
We use this method to show that there are an infinite number of
degenerate baryon states in the large-$N$ limit.
We also show that forward matrix elements of quark bilinear operators
satisfy the static quark-model relations in this limit, and enumerate the
corrections to these relations to all orders in $1/N$.
These results hold for any number of light quark flavors, and the methods
used can be extended to arbitrary operators.
Our results imply that for two flavors, the quark-model relations for the
axial currents and magnetic moments get corrections of order $1/N^2$.
For three or more flavors, the results are more complicated, and
corrections are generically of order $1/N$.
We write an explicit effective lagrangian which can be used to
carry out chiral perturbation theory calculations in the $1/N$ expansion.
Finally, we compare our results to what is expected from a chiral
constituent quark model.
\endabstract


\section{Introduction}

It is well-known that QCD has a non-trivial limit as $N\to\infty$, where
$N$ is the number of colors
\ref\largeN{G. 't Hooft, \NPB{72}{461}{1974};
\NPB{75}{461}{1974}.}.
This limit can be defined by writing the QCD covariant derivative as
\eq
D_\mu = \partial_\mu + \frac{i\mybar g}{\sqrt{N}} A_{\mu},
\eeq
where $A_\mu$ is the gluon field, and letting $N\to\infty$ holding
$\mybar g$ fixed.
QCD in the large-$N$ limit is believed to share many features
with QCD at $N = 3$, including confinement and chiral symmetry breaking.
(It can be shown that confinement implies chiral symmetry breaking
in the large-$N$ limit of QCD
\ref\CSB{S. Coleman and E. Witten, \PRL{45}{100}{1980}.}.)
Therefore, an expansion in $1/N$ around the large-$N$ limit of QCD can be
expected to give valuable insights into non-perturbative QCD phenomena.

The properties of baryons in the large-$N$ limit were first discussed in
detail by Witten \ref\Witten{E. Witten, \NPB{160}{57}{1979}.}, who argued
that baryons can be described by Hartree--Fock equations in this limit.
This approach gave a consistent picture of baryons, and showed how various
physical quantities scale with $N$.
Most of the subsequent work on baryons in the large-$N$ limit was done
in the context of the Skyrme model
\ref\Skyrme{T. H. R. Skyrme, \PRSL{A260}{127}{1961}.
For a modern review, see \eg\ I. Zahed and G. E. Brown, \PR{142}{1}{1986}.},
which has been argued to be exact as $N\to\infty$
\ref\WS{E. Witten, \NPB{223}{552}{1983}.}.
The baryon spectrum in the Skyrme model consists of an infinite number of
degenerate states.
It can also be shown that the axial current matrix elements obey the
static quark-model relations in the large-$N$ limit of the Skyrme model
\ref\bard{K. Bardakci, \NPB{243}{197}{1984}}.
This result was extended to all operators in
ref.\ \ref\manqm{A. V. Manohar, \NPB{248}{19}{1984}.}.
Another approach was initiated by Gervais and Sakita
\ref\GS{J.--L. Gervais and B. Sakita, \PRL{52}{87}{1984};
\PRD{30}{1795}{1984}.},
who showed that consistency of chiral perturbation theory with the
large-$N$ limit implies the quark-model relations for the axial
current matrix elements in the large-$N$ limit.
These methods were recently rediscovered and extended by Dashen and Manohar
\ref\DM{R. Dashen and A. V. Mahohar, \PLB{315}{425}{1993};
\PLB{315}{438}{1993}.}\
and Jenkins
\ref\Jenkins{E. Jenkins, \PLB{315}{441}{1993}.},
who considered $1/N$ corrections in this approach.

In this paper, we will use diagrammatic and group-theoretic methods to
analyze baryons in the $1/N$ expansion.
Using these methods, we derive the quark-model relations directly from
QCD in a way which makes clear the role of quarks as the constituents
of baryons.
These methods allow us to enumerate the corrections to these
relations to all orders in the $1/N$ expansion for the baryon masses and
matrix elements of quark bilinear operators for any number of light
quark flavors.
(The methods used can be easily extended to general matrix elements.)
For 2 flavors, our results imply that the quark-model relations for
the axial currents and magnetic moments get corrections of order $1/N^2$.
(For the non-singlet axial currents, this result was obtained in
ref.\ \DM.)
For 3 or more flavors, the results are more complicated, since the
baryon representations depend on $N$ in this case.
We find that the corrections to the quark-model relations are generically
of order $1/N$.

Combining these results with current algebra ideas, we then write an effective
lagrangian which can be used to carry out a simultaneous chiral and $1/N$
expansion for processes involving baryons and soft pseudo Nambu--Goldstone
bosons.
This lagrangian contains couplings that grow with $N$, so that it is not
manifest that its predictions are physically sensible.
We have verified the consistency of the predictions of this lagrangian
in some non-trivial cases (including all cases considered in
refs.\ \GS\DM\Jenkins), but we have not proved it in general.
However, we argue that this effective lagrangian is the most general one
with the right field content and symmetry structure in the large-$N$ limit,
and we therefore believe that it must be consistent on general grounds.

We conclude with some remarks comparing our results to what is expected
from a non-relativistic constituent quark model.

\section{$N$-Counting Rules for Baryons}

In this section, we derive our main results on the structure of baryon
masses and matrix elements in the $1/N$ expansion.
We begin by giving a rough idea of the nature of our argument.
Consider a baryon matrix element of the form
$\bra{\scr B'} T\hat{\scr O}_1 \cdots \hat{\scr O}_n \ket{\scr B}$.
The operators $\hat{\scr O}_1, \ldots, \hat{\scr O}_n$ can change the spin
and flavor of the quarks in the initial baryon consistent with
invariance under the flavor symmetry $SU(N_F)$.
If the velocities of the initial and final baryon states are equal, then
the spin and flavor changes must also be invariant under $SU(2)_v$, the
group of spatial rotations in the common baryon rest frame.
It turns out that invariance under the combined \sf\ symmetry is
especially restrictive when $N$ is large for states with total
spin $J \ll N$: the amplitude to change the spin or
flavor of a quark is suppressed by $1/N$ for such states.
This combinatoric result is responsible for the appearance of the static
quark-model relations in the large-$N$ limit for forward baryon matrix
elements.

To make this argument precise, our strategy is to write a diagrammatic
expansion for the quantities we wish to study, and then find properties of
this expansion that hold to all orders in $\mybar g$.
This method is not strictly rigorous, since it is not known whether
the quantities in question can actually be obtained by ``summing all
diagrams.''
However, the results derived here depend only on the $N$-counting of
the diagrammatic expansion, which we believe to be reliable.

We consider the QCD states
\eq
\label\thestate
\bk \equiv \scr B^{a_1 \al_1 \cdots a_N \al_N}
\epsilon^{A_1 \cdots A_N}
\hat a^\dagger_{a_1 \al_1 A_1} \cdots
\hat a^\dagger_{a_N \al_N A_N} \ket{0},
\eeq
where $a_1, \ldots, a_N = 1, \ldots, N_F$ are flavor indices,
$\al_1, \ldots, \al_N = \uparrow, \downarrow$ are spin indices,
and $A_1, \ldots, A_N = 1, \ldots, N$ are color indices.
The operators $\hat a^\dagger$ in eq.\ \thestate\ create a quark with
definite flavor, spin, and color, in a perturbative 1-quark state
$\ket\phi$.
We take $\ket\phi$ to have average momentum zero and momentum uncertainty
of order $\Lam_{\rm QCD}$.
Our results are insensitive to the precise nature of the state $\ket\phi$.
The important feature of the state $\bk$ for our analysis is that it has
the right quantum numbers to be a 1-baryon state at rest: it is a color
singlet, has unit baryon number, and has angular momentum and flavor
quantum numbers determined by the tensor $\scr B$.

The tensor $\scr B$ in eq.\ \thestate\ is symmetric under combined interchange
of spin and flavor indices, $a_1 \al_1 \leftrightarrow a_2 \al_2$, \etc.
$\scr B$ may be thought of as a quark-model wave function describing states
with spin $J = \frac 12, \ldots, \frac 12 N$ (for $N$ odd).
The Young tableaux of the $SU(N_F)$ representation with spin
$J$ is shown in fig.\ (1a).
We will {\it assume} that the lowest-lying baryon states in QCD have these
quantum numbers for arbitrarily large $N$, with the masses of the baryons
increasing with $J$.
This assumption is supported by the Skyrme model, and of course is true
for $N = 3$.

Under time evolution, the state $\bk$ projects onto the state
$\ket{\scr B}$, where $\ket{\scr B}$ is the minimal energy eigenstate of
the {\it full} hamiltonian with a nonzero overlap with
$\bk$:\footnote{$^\dagger$}
{For notational simplicity, we work in finite volume;
the spectrum of states is then discrete, and we take all states to be
normalized to unity.
If $\ket{\scr B}$ is a 1-baryon state with zero momentum, the overlap
$\braket{\scr B}{\scr B_0}$ goes to zero in the infinite volume limit
as $1 / \sqrt{V}$.
Because all of our results are independent of $\braket{\scr B}{\scr B_0}$,
this is harmless.}
\eq
\eqalign{
\label\GML
\hat U(0, -T) \bk &\too \ket{\scr B} \braket{\scr B}{\scr B_0}
e^{-iE_{\scr B} T} + \cdots, \cr
\bb \hat U(T, 0) &\too \braket{\scr B_0}{\scr B} \bra{\scr B}
e^{-iE_{\scr B} T} + \cdots, \cr}
\eeq
as $T \to -i\infty$, where the ellipses denote terms which are exponentially
suppressed.
(The analytic continuation is the usual Wick rotation, which is justified
by the assumption that the only singularities of the time-evolution operator
are those required by unitarity and crossing.)
Eq.\ \GML\ allows us to write matrix elements involving the state
$\ket{\scr B}$ as a sum of diagrams with external states of the form $\bk$.
In our approach the state $\bk$ plays a role similar to that of the
{\it perturbative} vacuum, where a formula like eq.\ \GML\ is used to derive
a diagrammatic expansion for time-ordered products in the {\it interacting}
vacuum.

If we choose the tensor $\scr B$ in eq.\ \thestate\ to have definite
spin--flavor quantum numbers, then by the assumptions above, there is
a 1-baryon state with the same quantum numbers as $\bk$.
We would like to identify $\ket{\scr B}$ with this 1-baryon state, but we
must ensure that there is no state with lower energy with the
same quantum numbers.
Since $\ket{\scr B}$ has unit baryon number, the candidates for this state
contain a single baryon and any number (possibly zero) of mesons,
glueballs, and exotics.
For example, if $\ket{\scr B}$ has spin $J$, we can consider states
consisting of a spin-$(J - 1)$ baryon and a meson.
Since mesons have masses which are order 1 in the $1/N$ expansion, this
state will have more energy than the 1-baryon state provided that
$M_J - M_{J - 1} \to 0$ as $N \to \infty$.
Generalizing these considerations, we see that the method we are using
allows us to study those states which become degenerate with the
lowest-lying baryon states (assumed to be the $J = \frac 12$ multiplet)
in the large-$N$ limit.\footnote{$^\dagger$}
{One might worry about states containing a baryon and soft pions in
the chiral limit.
However, we can add $SU(N_F)$-preserving current quark mass term to make
the pions massive and remove the degeneracy of these states
with the one-baryon states.
The smoothness of the chiral limit then allows us to extrapolate our
results to zero current quark masses.}
We already know from previous work that there are an infinite number
of such states in the large-$N$ limit;
this result will also emerge self-consistently from our analysis.

\subsection{Masses}

In this subsection, we derive our main results on the baryon masses.
We will work in the limit where the current quark mass differences and
electromagnetism are turned off, so that the theory has an exact $SU(N_F)$
flavor symmetry.
In this limit, the baryon spectrum consists of multiplets with spin
$J = \frac 12, \ldots, \frac 12 N$ with the baryons degenerate in each
multiplet.
(We will show later how to treat $SU(N_F)$ breaking perturbatively
using chiral perturbation theory.)

Consider the quantity
\eq
\label\zdef
Z \equiv \bb e^{-i\hat HT} \bk,
\eeq
where $\hat H$ is the QCD hamiltonian.
Using eq.\ \GML, we have
\eq
\label\energy
\frac 1T \ln Z \too -iE_{\scr B},
\eeq
as $T\to -i\infty$.
To define a diagrammatic expansion for $Z$, we write the hamiltonian as
$\hat H = \hat H_0 + \del\hat H$.
We choose $\hat H_0$ to be a 1-body operator such that $\bk$ is an eigenstate;
consequently, $\del\hat H$ will in general contain 1-body ``interactions,'' but
these do not affect the $N$-counting of the expansion.
Once again, our results are independent of the precise form of $\hat H_0$ and
$\del \hat H$.
We then write
\eq
\label\graff
Z = e^{-iE_0 T} \bb \hat U_I(\sfrac 12 T, -\sfrac 12 T) \bk,
\eeq
where $\hat H_0 \bk = E_0 \bk$, and
\eq
\hat U_I(t_1, t_0) \equiv
T\exp\left[-i\int_{t_0}^{t_1} dt\, \del\hat H_I(t) \right], \qquad
\hat H_I(t) \equiv e^{i\hat H_0 t} \del\hat H e^{-i\hat H_0 t},
\eeq
Eq.\ \graff\ can be expanded as a sum of Feynman diagrams using a simple
adaptation of standard techniques.
Each term in the expansion of eq.\ \graff\ in powers of $\del\hat H$
consists of a time-ordered product of operators.
Using Wick's theorem, this can be written as a sum of all possible normal
orderings and contractions.
The terms in which all operators are contracted are just the usual
``vacuum'' graphs which sum to give the vacuum energy.
However, because the state $\bk$ contains particles, the normal-ordered
terms can also contribute.
We represent these contributions diagrammatically by associating a
dashed line for each external quark creation or annihilation operator, as
shown in fig.\ (2b,c).
In each such term, it is understood that all creation and annihilation
operators are normal ordered.

In this diagrammatic expansion, the mass of the state $\ket{\scr B}$ is
given by
\eq
\label\opex
M_{\scr B} \equiv E_{\scr B} - E_{\rm vac} =
\sum_r c_r \bb \hat{\scr O}^{(r)} \bk,
\eeq
where $E_{\rm vac}$ is the vacuum energy and
$\hat{\scr O}^{(r)}$ is an $r$-body operator of the form
\eq
\label\QCDop
\hat{\scr O}^{(r)} =
X^{a_1 \al_1 \cdots a_r \al_r}_{b_1 \beta_1 \cdots b_r \beta_r}
\hat a^\dagger_{a_1 \al_1 A_1} \hat a^{b_1 \beta_1 A_1} \cdots
\hat a^\dagger_{a_r \al_r A_r} \hat a^{b_r \beta_r A_r}.
\eeq
The external creation and annihilation operators again create a quark
with definite spin, flavor, and color, in the 1-quark state $\ket\phi$ used
to define the state $\bk$.
(We can take $\ket\phi$ to be one of a complete basis of 1-quark
states, and write all contributions in terms of creation and annihilation
operators in this basis.
Operators which annihilate quarks in 1-quark states other than $\ket\phi$
then vanish on states of the form $\bk$.)
To see that eq.\ \QCDop\ is the most general form of a color-singlet
$r$-body operator, note that operators involving color $\ep$ symbols can
always be reduced to this form by using the identity
\eq
\label\epid
\ep^{A_1 \cdots A_N} \ep_{B_1 \cdots B_N} = \det\Del,
\quad{\rm where}\quad
\Del_{st} \equiv \del^{A_s}_{B_t}
\quad (s,t = 1, \ldots, N).
\eeq

Since QCD conserves angular momentum and flavor, the operators
$\hat{\scr O}^{(r)}$ in eq.\ \opex\ must be singlets under \sf, where
$SU(2)_v$ denotes the group of spatial rotations in the baryon rest
frame.
The classification of these operators is a group-theoretical problem which
is solved in appendix A.
Since the results concern only the spin and flavor structure, it is
convenient to express the results in terms of a spin--flavor Fock space:
we define
\eq
\rket{\scr B} \equiv \scr B^{a_1 \al_1 \cdots a_N \al_N}
\al^\dagger_{a_1 \al_1} \cdots \al^\dagger_{a_N \al_N} \rket 0.
\eeq
Because the baryon wavefunction is symmetric under
$a_1 \al_1 \leftrightarrow a_2 \al_2$ \etc., the creation operators
$\al^\dagger$ are bosonic.
We use the ``curved bra--ket'' notation to distinguish these states from
QCD states.
Given any QCD operator $\hat{\scr O}$ appearing in eq.\ \opex, we can
construct an operator in the spin--flavor Fock space such that
$\bb \hat{\scr O} \bk \propto \rbra{\scr B} \scr O \rket{\scr B}$, so
we can write
\eq
\label\opextoo
M_{\scr B} \equiv E_{\scr B} - E_{\rm vac} =
\sum_r c_r \rbra{\scr B} \scr O^{(r)} \rket{\scr B},
\eeq
More details are given in the appendix.

The result of the classification is that an arbitrary $r$-body operator
appearing in eq.\ \opextoo\ can be written as a linear combination of
\eq
\op{1}^{r - 2s} \op{J^2}^s.
\eeq
Here $\op 1$ is the quark number operator and $\op{J^2}$ is the usual
angular momentum Casimir operator:
\eq
\op 1 \rket J = N \rket J, \qquad
\op{J^2} \rket J = J(J + 1) \rket J,
\eeq
where $\rket J$ denotes any state $\rket{\scr B}$ with total spin $J$.
We emphasize that this result holds for any number of flavors $N_F \ge 2$.
This result can be understood heuristically by noting that $\op{J^2}$ is
a Casimir operator which suffices to label the possible representations of
$\rket{\scr B}$, since the flavor representations are determined by
$J$ (see fig.\ (1a)).
Of course, we can eliminate the operator $\op 1$ using the relation
$\op 1 = N$ which holds on states with baryon number 1.

One way to determine the $N$ dependence of the coefficients in
eq.\ \opex\ is to use the fact that the baryon mass is order $N$ \Witten.
Because we assume that the baryon masses increase with $J$, $M_{\scr B}$
must be greater than the mass of the $J = \frac 12$ baryon multiplet, which
is of order $N$.
On the other hand, $M_{\scr B}$ can be no greater than the mass of the
1-baryon state with the same quantum numbers, which is also of order $N$.
Therefore, $M_\scr B \sim N$ for all states $\ket{\scr B}$, and each term in
eq.\ \opex\ must be at most of order $N$.
This implies
\eq
\label\mbres
M_{\scr B} = M_0 + \sum_s c_s \rbra{\scr B} \op{J^2}^s \rket{\scr B},
\qquad c_s \lsim \frac 1{N^{2s - 1}} + \cdots,
\eeq
where $M_0 \sim N$ and the ellipses denote terms higher order in $1/N$.
Since there seems to be no reason to suppose otherwise, we will assume
that the inequalities in eq.\ \mbres\ are saturated.

Eq.\ \mbres\ is our main result for the baryon masses.
We can immediately see that the masses of states with $J \sim 1$ become
degenerate in the large-$N$ limit.
This allows us to conclude that the states with $J \sim 1$ are in fact
1-baryon states, as discussed above.
The fact that the baryon spectrum consists of an infinite tower of degenerate
states in the large-$N$ limit is a well-known result of the Skyrme model,
and is also a consequence of the methods of ref.\ \GS.
We also see that the leading mass differences between states with
$J \sim 1$ are given by
\eq
\label\massdiff
M_J - M_{J'} = \frac\mu N \left[ J(J+1) - J'(J'+1) \right].
\eeq
This result was derived recently in ref.\ \Jenkins\ by demanding consistency
of chiral perturbation theory in the large-$N$ limit.
We have extended these results by classifying the corrections to all orders
in $1/N$.
A qualitative picture of the baryon spectrum for large values of $N$
is shown in fig.\ (1b).

The previous discussion relied on the results of ref.\ \Witten, and does not
make use of the full power of the diagrammatic approach.
We now consider the diagrammatic expansion in more detail and show that
eq.\ \mbres\ can be derived entirely within this approach.

We begin by showing that $\ln Z$ can be expressed as a sum of connected
diagrams.
The expansion of $Z$ clearly includes disconnected diagrams such as the
ones shown in fig.\ (2c).
A diagram with $n$ disconnected components is proportional to $T^n$, where
$T$ is the time extent.
This can be most easily seen in position space where there is one overall
time integration for each disconnected component ranging from $-\frac 12 T$
to $\frac 12 T$.
However, since $\ln Z = -iE_{\scr B}T$, we know that the disconnected
contributions must exponentiate in some way.
For the vacuum graphs (\ie\ those with no external creation or annihilation
operators), this exponentiation is very simple:
it is a standard result in quantum field theory that the disconnected
diagrams have the right combinatoric factors to combine into an exponential.
The graphs with external creation and annihilation operators have
the same combinatoric factors, so we can write
\eq
\ln Z = -iE_{\rm vac} T + \ln \bb \no{e^\Del} \bk,
\eeq
where $E_{\rm vac}$ is the sum of all connected vacuum graphs, and $\Del$
is the sum of connected graphs with external creation and annihilation
operators.
If $\bk$ is a state with definite total spin, then it is an eigenstate of
any \sf\ singlet operator of the form of eq.\ \QCDop, so that
\eqa
\ln \bb \no{e^\Del} \bk &= \bb \ln\no{e^\Del} \bk \eolnn
\label\opexp
&= \bb\no\Del\bk +
\sfrac 12 \bb\no{\Del^2} - (\no\Del)^2 \bk + \cdots \eeol
\eeq
We see that the disconnected diagrams with external creation and
annihilation operators do not exponentiate directly.

However, it is not hard to see that the \rhs\ of eq.\ \opexp\ is a sum
of connected diagrams of a new type.
Consider the $O(\Del^n)$ terms in the expansion of $\ln Z$.
By anticommuting the quark operators, each $O(\Del^n)$ term can be written
in a standard form as $(\no\Del)^n + $ terms with anticommutators.
The relevant anticommutator is
$S_H(x - y) \equiv \{ \psi^-(x), \mybar\psi^+(y) \}$, where $\psi$ is
the quark field and we use $\pm$ to indicate the creation/annihilation
part of the field.
$S_H$ is a Green's function, and is denoted diagrammatically by a dashed
line.
(Since it appears only in internal lines, there can be no confusion with the
external creation and annihilation operators.)
Physically, $S_H$ is a correction to the propagator required to properly
describe the propagation of ``holes'' (the absence of a
quark in the initial state).
A typical diagram involving hole propagators is shown in fig.\ (2d).
We now say that a diagram in this standard form is ``connected'' if it is
connected by either Feynman propagators or hole propagators.
A graph with $n$ connected components is proportional to $T^n$, so only
the connected graphs can contribute to $\ln Z$;
the disconnected diagrams must cancel.
It is easy to see explicitly that this happens in the $O(\Del^2)$ term
in eq.\ \opexp.
This expansion in terms of connected diagrams with hole propagators is a
generalization of the connected cluster expansion for Hartree--Fock
perturbation theory which is well known in many-body physics
\ref\Goldstone{J. Goldstone, \PRSL{A239}{267}{1957}.}.

The $N$-counting of diagrams is now very simple:
we associate one power of $1/\sqrt N$ to each gluon vertex and one power
of $N$ to each internal color loop.
(Color loops are most easily counted using the double-line notation
\largeN.)
In this way, one can see that the diagrammatic expansion for the mass of
the state $\ket{\scr B}$ can be written
\eq
\label\ncount
M_{\scr B} = \sum_r c_r \rbra{\scr B} \scr O^{(r)} \rket{\scr B}, \qquad
c_r \lsim \frac 1{N^{r - 1}},
\eeq
where $\scr O^{(r)}$ is an $r$-body operator in the spin--flavor Fock space.
Note that
\eqa
\op 1 &= \al^\dagger_{a\al} \al^{a\al}
= \hbox{\rm 1-body\ operator}, \eol
\op{J^2} &= \op{J_j} \op{J_j}
= \hbox {\rm 2-body\ operator}, \eeol
\eeq
where $\op{J_j} \equiv \al^\dagger_{a\al} (J_j)^\al_\beta \al^{a\beta}$.
Eq.\ \ncount\ therefore gives rise to the results for the masses
quoted above.

\subsection{Matrix Elements}

We now extend the techniques described above to operator matrix elements.
Using eq.\ \GML, we can write\footnote{$^\dagger$}
{We choose the phases of $\ket{\scr B}$ and $\ket{\scr B'}$ so that
$\braket{\scr B_0}{\scr B}$ and $\braket{\scr B'_0}{\scr B'}$ are real.}
\eq
\label\transeq
\bra{\scr B'} T \hat{\scr O}_1(t_1) \cdots \hat{\scr O}_n(t_n)
\ket{\scr B} =
\frac{\bpb T \hat{\scr O}_{I1}(t_1) \cdots
\hat{\scr O}_{In}(t_n) \hat U_I \bk}
{\bpb \hat U_I \bpk^{1/2} \,
\bra{\scr B_0} \hat U_I \ket{\scr B_0}^{1/2}}.
\eeq
Here, $\hat{\scr O}_1, \ldots, \hat{\scr O}_n$ are Heisenberg-picture
operators, $\hat{\scr O}_{I1}, \ldots, \hat{\scr O}_{In}$ are the
corresponding interaction-picture operators, and we use the abbreviation
$\hat U_I \equiv \hat U_I(\frac 12 T, -\frac 12 T)$.
The state $\bpk$ has the same form as eq.\ \thestate, possibly with
different spin and flavor quantum numbers.
(In particular, it is defined in terms of the same 1-quark state $\ket\phi$
used to define $\bk$.)
The times $t_1, \ldots, t_n$ in eq.\ \transeq\ must be lie within a fixed
finite time interval as $T\to\infty$ in order for the states $\bk$ and
$\bpk$ to project onto the corresponding interacting states $\ket{\scr B}$
and $\ket{\scr B'}$.

We now consider the diagrammatic expansion for the \rhs\ of eq.\ \transeq.
The numerator can be written as the sum of diagrams with insertions of the
operators $\scr O_1, \ldots, \scr O_n$.
The denominator can be written as a sum of diagrams with combinatoric
factors obtained by expanding out the square roots.
The numerator and denominator can be combined into a sum of diagrams
involving hole propagators in a manner similar to that of eq.\ \opexp.
The diagrammatic expansion therefore gives
\eq
\label\matexp
\bra{\scr B'} T \hat{\scr O}_1(\vec p_1, t_1) \cdots
\hat{\scr O}_n(\vec p_n, t_n) \ket{\scr B} =
\sum_\ell F_\ell(\vec p_1, t_1, \ldots, \vec p_n, t_n)
\bb \hat{\scr O}_\ell^{(r_\ell)} \bk,
\eeq
where we have explicitly indicated the dependence on the 3-momenta of the
operators.
Each term in this series can be thought of as a form factor.
The coefficient $F_\ell$ contains all of the kinematic information of the
form factor.
$\hat{\scr O}_\ell^{(r_\ell)}$ is an $r_\ell$-body operator of the form
of eq.\ \QCDop\ (with no kinematic dependence) which gives the spin and
flavor dependence of the form factor.

We now argue that only connected diagrams contribute to the \rhs\ of
eq.\ \matexp.
First, we note that we can discard any graph in which the insertions of
the operators $\hat{\scr O}_1, \ldots \hat{\scr O}_n$ are in different
connected components by giving the operators definite 3-momenta and
restricting attention to ``unexceptional'' kinematic regions where no
proper subset of the 3-momenta sum to zero.
A graph with $n$ connected components is therefore  proportional to
$T^{n - 1}$, since the overall time integrations for the $n - 1$ connected
components which do not contain the operator insertions can range from
$-\frac 12 T$ to $\sfrac 12 T$;
the connected component containing the operators
has no overall time integral because the operators act at definite times.
However, because
$\bra{\scr B'} T \hat{\scr O}_1(t_1) \cdots \hat{\scr O}_n(t_n)
\ket{\scr B} \sim T^0$,
only the connected diagrams contribute.

By counting powers of $N$ in connected graphs, we find that the coefficient
of an $r_\ell$-body operator in eq.\ \matexp\ is
\eq
\label\opncount
F_\ell \lsim \frac 1{N^{r_\ell + L - 1}},
\eeq
where $L$ is the minimum number of quark loops in graphs that contribute
to the matrix element.
These results are valid as long as the momenta of the
operators are held fixed as $N\to\infty$.
In this case, the initial and final baryon velocity are the same
in the large-$N$ limit.
If we attempt to describe processes in which the baryon velocity
changes, some of the momenta of the operators must become large with
$N$.
In this case, there is additional $N$ dependence coming from the
overlap integrals between the initial and final states and the matrix
elements are suppressed by $e^{-N}$ \Witten.

As was the case for the baryon masses, we can now obtain interesting
results from a group-theoretic classification of the operators that
appear on the \rhs\ of eq.\ \matexp.
In terms of the spin--flavor Fock space, we have
\eq
\label\matexptoo
\bra{\scr B'} T \hat{\scr O}_1(\vec p_1, t_1) \cdots
\hat{\scr O}_n(\vec p_n, t_n) \ket{\scr B} =
\sum_\ell F_\ell(\vec p_1, t_1, \ldots, \vec p_n, t_n)
\rbra{\scr B} \scr O_\ell^{(r_\ell)} \rket{\scr B}.
\eeq
Note that the coefficients $F_\ell$ may have $SU(2)_v$ indices coming from
non-trivial dependence on the angles of the momenta
$\vec p_1, \ldots, \vec p_n$, so $\scr O_\ell$ does not necessarily have
the same quantum numbers as the product of operators on the \lhs.

The matrix elements of most interest are forward matrix elements of a single
quark bilinear of the form $\mybar\psi \Gam W \psi$, where $\psi$ is
the quark field, $W$ is a flavor matrix, and $\Gam$ is a Dirac spinor
matrix.
For such matrix elements, the operators $\scr O_\ell$ transform under
\sf\ the same way as the tensor $W^a_b \Gam^\al_\beta$.
Here, $\Gam^\al_\beta$ is the static projection of $\Gam$, \ie
\eq
(\gam_j \gam_5)^\al_\beta = (J_j)^\al_\beta, \quad
(\gam_0)^\al_\beta = \del^\al_\beta,
\eeq
\etc.
As shown in appendix A, the most general operator with these quantum
numbers is a combination of
\eq
\label\opcoll
\op{W \Gam}, \quad
\tr(W) \op{\Gam}, \quad
\op{W} \op{\Gam}, \quad
\op{W J_j} \op{J_j} \op{\Gam},
\eeq
multiplied by powers of $\op{J^2}$ and $\op 1$.
Here,
\eq
\op{W \Gam} \equiv \al^\dagger_{a\al} W^a_b \Gam^\al_\beta
\al^{b\beta},
\eeq
\etc.
This classification holds for any number of flavors $N_F \ge 2$.
The operator $\op{W \Gam}$ gives rise to the quark model predictions
for the matrix elements, which obey the well-known $SU(2N_F)$
``spin--flavor'' relations.
The methods used in the appendix for classifying operators can be easily
extended to arbitrary operators.

The $N$ dependence of the coefficients of the operators in
eq.\ \opcoll\ can be read off from the rule in eq.\ \opncount.
The operator $\tr(W) \op{\Gam}$ has contributions from diagrams
involving at least one quark loop, and so has coefficient $1/N$.
The other operators do not require quark loops, and so have coefficients
$1/N^{r - 1}$ for an $r$-body operator.

For $N_F > 2$, the $N$-dependence of the matrix elements is somewhat
complicated due to the fact that the size of the $SU(2N_F)$ representation
with given $J^2$ grows with $N$ (see fig 1a).
We will therefore begin by discussing the case $N_F = 2$, where the
baryon states are characterized by $I = J = \frac 12, \frac 32, \ldots$.
In this case, direct calculation shows that
\eq
\label\redund
\op{W J_j} \op{J_j} = \frac{N + 2}4 \op{W},
\eeq
for $\tr W = 0$.
Therefore, the last operator in eq.\ \opcoll\ is
redundant.\footnote{$^\dagger$}{We thank A. V. Manohar for
pointing out this redundancy, which was missed in the initial
version of this work.
We also thank A. Cohen for discussions on this point, and on the
$N$-dependence of amplitudes involving baryons with finite strangeness
in the 3-flavor case.}
(This result does not hold for $N_F > 2$.)

We now consider the $N$ dependence of matrix elements of the
operators in states with $J \sim 1$.
If both $W$ and $\Gam$ are equal to the unit matrix (\eg\ for the flavor
singlet scalar density), then the matrix element is clearly of order $N$.
If one of the matrices $W$ and $\Gam$ is the unit matrix and the other is
a traceless matrix (\eg\ for the flavor non-singlet vector charge), the
matrix element is of order 1, since $\op{W \Gam}$ is an
\sf\ generator in this case.
The only case which is not immediately obvious is when both $W$ and $\Gam$
are traceless, corresponding to the flavor non-singlet axial current.
Direct calculation shows that the matrix element is of order $N$ in this
case.

These results, along with the order of the leading corrections to the
large-$N$ limit, are shown in table 1.
Note that in every case, the quark-model result corresponding to the
operator $\op{W\Gam}$ is exact in the large-$N$ limit, and the corrections
start at order $1/N^2$.


\vbox{\vskip 20pt \centerline{
\vbox{\offinterlineskip
\halign{\vrule#
&\hfil#\hfil&\vrule#
&\hfil#\hfil&\vrule#
&\hfil#\hfil&\vrule#
&\hfil#\hfil&\vrule#
&\hfil#\hfil&\vrule#
\cr
\noalign{\hrule}
height2pt
&\omit&
&\omit&
&\omit&
&\omit&
&\omit&
\cr
& &
& \ $W, \Gam = 1$ &
& \ $\tr W = 0$, $\Gam = 1$ &
& \ $W = 1$, $\tr \Gam = 0$ &
& \ $\tr W, \tr \Gam = 0$ &
\cr
height2pt
&\omit&
&\omit&
&\omit&
&\omit&
&\omit&
\cr
%
%
\noalign{\hrule}
height2pt
&\omit&
&\omit&
&\omit&
&\omit&
&\omit&
\cr
& $\op{W \Gam}$ &
& $N$ &
& $1$ &
& $1$ &
& $N$ &
\cr
height2pt
&\omit&
&\omit&
&\omit&
&\omit&
&\omit&
\cr
\noalign{\hrule}
height2pt
&\omit&
&\omit&
&\omit&
&\omit&
&\omit&
\cr
& $\tr(W) \op{\Gam} / N$ &
& $1^*$ &
& $0$ &
& ${1/N}^*$ &
& $0$ &
\cr
height2pt
&\omit&
&\omit&
&\omit&
&\omit&
&\omit&
\cr
\noalign{\hrule}
height2pt
&\omit&
&\omit&
&\omit&
&\omit&
&\omit&
\cr
& $\op{W} \op{\Gam} / N + \hc$ &
& $N^*$ &
& $1^*$ &
& $1^*$ &
& $1/N$ &
\cr
height2pt
&\omit&
&\omit&
&\omit&
&\omit&
&\omit&
\cr
\ignore
\noalign{\hrule}
height2pt
&\omit&
&\omit&
&\omit&
&\omit&
&\omit&
\cr
& \ $\op{W J_j}\op{J_j}\op{\Gam}/N^2 + \hc$ &
& $1/N$ &
& $1$ &
& $1/N^2$ &
& $1/N$ &
\cr
height2pt
&\omit&
&\omit&
&\omit&
&\omit&
&\omit&
\cr
\endignore
\noalign{\hrule}
height2pt
&\omit&
&\omit&
&\omit&
&\omit&
&\omit&
\cr
& \ $\op{W \Gam}\op{J^2}/N^2 + \hc$ &
& $1/N$ &
& $1/N^2$ &
& $1/N^2$ &
& $1/N$ &
\cr
height2pt
&\omit&
&\omit&
&\omit&
&\omit&
&\omit&
\cr
\noalign{\hrule}
}}}\vskip 10pt }
{\leftskip=30pt\rightskip=30pt\noindent
Table 1. Leading corrections to quark bilinear matrix elements for 2
light quark flavors.
The corrections marked with a $*$ are proportional to the lowest-order
values.
\par}\vskip 10pt

We now consider briefly the extension of these results to the case
$N_F = 3$.
In this case, the $SU(N_F)$ representations corresponding to spin
$J \sim 1$ contain $\sim N$ states.
We identify the physical baryon states with the large-$N$ states that have
the same isospin and strangeness.
The $N$-dependence of the matrix elements is now more complicated due to
the dependence on the strangeness $S$ of the baryon states.
For example, the $\Del S = 1$ semileptonic hyperon decays proceed through
matrix elements of the QCD operator
\eq
\hat{\scr O}^\mu_{\Del S = 1} = \bar\psi_L \gam^\mu X \psi_L,
\eeq
where
\eq
\psi = \pmatrix{u \cr d \cr s \cr}, \qquad
X = \pmatrix{0 & 0 & 1 \cr 0 & 0 & 0 \cr 0 & 0 & 0 \cr}.
\eeq
In this case, the corresponding 1-body operator in the spin--flavor Fock
space is $\op{X J_j}$.
This has matrix elements of order $\sqrt{|S|N}$ between states with strangeness
$S$, since it removes a quark in a state with occupation number
$\sim S$ and places it into a state with occupation number $\sim N$.
The computation of these matrix elements is rather complicated, and
must be handled on a case-by-case basis.
Generically, we find that corrections to the quark-model relations which
depend on $SU(3)_F$ symmetry (and not only isospin symmetry alone) are of
order $1/N$.
These issues will be discussed more fully in a future work, where we
will also consider in detail whether the $1/N$ expansion is
quantitatively relevant for $N = 3$.

We close this section with some general observations on these results.
The quark-model relations we have derived in the large-$N$ limit are often
said to be the consequence of an $SU(2N_F)$ spin--flavor symmetry.
It is therefore natural to ask whether there is any sense in which QCD
posseses such a symmetry in the large-$N$ limit.
Because any $SU(2N_F)$ symmetry should contain the $SU(N_F)$ flavor
symmetry of the light quarks, {\it all} particles with $SU(N_F)$ quantum
numbers must transform non-trivially under $SU(2N_F)$.\footnote{$^\dagger$}
{If all particles with $SU(N_F)$ quantum numbers decoupled from the
baryons in the large-$N$ limit, then the $SU(2N_F)$ symmetry could arise
as a superselection rule in this limit.
However, this does not happen.
For example, the pion--nucleon scattering amplitude is order 1 in the
$1/N$ expansion (see eq.\ (46) below).}
(This is in contrast with heavy-quark symmetry, which commutes with
the chiral symmetry of the light quark flavors.)
However, it is clear that the observed mesons (for example) do not lie
in $SU(6)$ multiplets.
Because of this difficulty, we do not know of any sense in which the
quark-model relations are the manifestation of an extra symmetry in the
large-$N$ limit.

Finally, we note that it may seem paradoxical that ``non-relativistic''
relations such as the  quark-model relations can arise in a
relativistic system of quarks and gluons such as the one we have been
studying.
However, the quark-model relations we have derived are non-relativistic
only in the sense that Lorentz invariance implies that they can hold in
at most one reference frame.
In our approach, this frame is picked out by the baryon's velocity,
and the resulting picture is completely Lorentz covariant.

\vfill\eject
\section{Chiral Perturbation Theory}

Using the results for operator matrix elements derived above, we can
write down an effective lagrangian which describes the
interactions of baryons with soft pions in the $1/N$ expansion.

We begin by discussing the baryons.
Because the baryon mass is of order $N$, we can describe the baryons using
a heavy-particle effective field theory
\ref\heavyB{E. Jenkins and A. V. Manohar, \PLB{255}{558}{1991}.}.
We write the baryon momentum as
\eq
\label\heavymom
P = M_0 v + k,
\eeq
where $M_0 \sim N$ is a baryon mass and $v$ is a 4-velocity ($v^2 = 1$)
which defines the baryon rest frame.
We then write an effective field theory in terms of baryon fields whose
momentum modes are the residual momenta $k$.
The reason that this effective theory is useful is that all kinematic
dependence on $M_0$ is removed.

The Lorentz structure also simplifies considerably in a heavy-particle
effective theory.
All Lorentz invariants in the spin-$\frac 12$ representation can be
constructed from $v^\mu$, $\ep^{\mu\nu\lam\rho}$, and the spin matrix
\eq
s^\mu \equiv \sfrac 12 (\gam^\mu - v^\mu \sla v)\gam_5.
\eeq
We find it convenient to simply work in the rest frame defined by $v$
and use a non-relativistic notation.

To describe the baryons with spin $\frac 12, \frac 32, \ldots$,
we use the baryon field $\scr B^{a_1 \al_1 \cdots a_N \al_N}(x)$.
(In a relativistic notation, these fields would satisfy $N$ constraints of
the form $P_+ \scr B = 0$, where the particle projection operator
$P_+ \equiv \frac 12 (1 + \sla v)$ can act on any of the spinor indices.)
The heavy field $\scr B$ is defined using a {\it common} mass $M_0$ for all
of the baryons, which we choose to be the mass of the $J = \frac 12$
multiplet in the $SU(N_F)$ limit.
With this choice, $M_0$ is close to the physical mass of baryons with
$J \sim 1$.

The large-$N$ relations derived in the previous section hold only
for baryons with $J \sim 1$, while the effective lagrangian we are
describing also includes baryon fields with $J \sim N$.
Because these states have masses $M_0 + O(N)$, their
effects on the dynamics of the states with $J \sim 1$ will be local (on the
scale of external momenta which are order 1 in the $1/N$ expansion)
and suppressed by powers of $1/N$.
(We do not want to consider an effective theory with these
states integrated out, since integrating out a subset of the baryon
states violates the spin--flavor relations explicitly.)
Furthermore, baryons with $J \sim N$ will only appear in intermediate
states if we consider the insertion of operators with spin $\sim N$ or
if we carry consider diagrams involving $\sim N$ vertices.
However, it is still important to know whether these contributions, when
they appear, signal a breakdown of the $1/N$ expansion as defined by this
effective lagrangian.
If the contributions from baryons with $J \sim N$ have the same form as
the tree-level terms, then these contributions are harmless.
While we believe this to be the case, we have no formal proof.
We will see below that the consistency of the effective lagrangian we
are describing is non-trivial even at low orders in the $1/N$ expansion.

We now discuss the light mesons.
The main new feature of chiral dynamics in the large-$N$ limit is that
the effects of the axial anomaly are subleading in $1/N$, so that the
pattern of spontaneous symmetry breakdown in the large-$N$ limit is
\ref\etaprime{E. Witten, \NPB{156}{269}{1979};
P. Di Vecchia, \PLB{85}{357}{1979};
G. Veneziano, \NPB{159}{213}{1979}.}\CSB
\eq
SU(N_F)_L \times U(1)_L \times SU(N_F)_R \times U(1)_R
\too SU(N_F)_{L + R} \times U(1)_{L + R}.
\eeq
Therefore, in addition to the usual $N_F^2 - 1$ Nambu--Goldstone bosons
(``pions''), there is an extra Nambu--Goldstone boson in the large-$N$
limit, which can be identified with the $\eta'$.
At higher orders in $1/N$, $U(1)_{L - R}$ is broken explicitly.
In this section, we restrict ourselves to making some remarks on
chiral perturbation theory which do not involve the $\eta'$ in an
essential way, so we will ignore the presence of the light $\eta'$ in the
following discussion.

The $N_F^2 - 1$ pions are described by a field
\eq
\xi(x) = e^{i\Pi(x) / f_\pi},
\eeq
which is taken to transform under $SU(N_F)_L \times SU(N_F)_R$ as
\eq
\xi \mapsto L \xi U^\dagger = U \xi R^\dagger,
\eeq
where this equation implicitly defines $U$ as a function of $L$, $R$,
and $\xi$.
The effective lagrangian is most conveniently written in terms of the
hermitian fields
\eq
V_\mu \equiv \frac i2\left(\xi \partial_\mu \xi^\dagger
+ \xi^\dagger \partial_\mu \xi\right), \qquad
A_\mu \equiv \frac i2\left(\xi \partial_\mu \xi^\dagger
-\xi^\dagger \partial_\mu \xi\right),
\eeq
transforming as
\eq
V_\mu \mapsto U V_\mu U^\dagger - i\partial_\mu U U^\dagger, \qquad
A_\mu \mapsto U A_\mu U^\dagger.
\eeq

With these definitions, it is easy to write down the effective lagrangian.
In order to keep track of the tensor algebra, it is convenient to once
again utilize the spin--flavor Fock space.
For example, we write the heavy field $\scr B$ in terms of
\eq
\rket{\scr B} \equiv \scr B^{a_1 \al_1 \cdots a_N \al_N}
\al^\dagger_{a_1 \al_1} \cdots \al^\dagger_{a_N \al_N} \rket 0.
\eeq
The chiral covariant derivative acting on baryon fields can then be
written as
\eq
\nabla_\mu \rket{\scr B} = \left( \partial_\mu - i\op{V_\mu} \right)
\rket{\scr B},
\eeq
and the first few terms in the effective lagrangian involving the baryon
fields are
\eq
\label\effL
\scr L_{\rm eff} = \rbra{\scr B} i\nabla_0 \rket{\scr B}
+ g \rbra{\scr B} \op{A\cdot s} \rket{\scr B}
+ c \rbra{\scr B} \op{m} \rket{\scr B} + \cdots,
\eeq
where $m$ is the quark mass matrix spurion, and
\eq
\op{A\cdot s} \equiv (s^j)^\al_\beta (A_j)^a_b
\al^\dagger_{a\al} \al^{b\beta}, \qquad
\op{m} \equiv m^a_b \al^\dagger_{a\al} \al^{b\al}.
\eeq
To find the $N$ dependence of the coefficients $g$ and $c$ in eq.\ \effL,
we use the fact that the terms in the effective lagrangian can be related
to matrix elements of QCD operators.
The term proportional to $g$ gives rise to a contribution to the matrix
element of the axial current, while the term term proportional to $c$ gives
rise to a contribution to the matrix element of the scalar density.
Comparing with eq.\ \opncount, we see that $g$ and $c$ are both of order 1
in the $1/N$ expansion.
In general, it is not hard to see that a general term in the effective
lagrangian involving the baryon fields will have the form
\eq
\label\effncount
\frac 1{N^{r - 1}} \rbra{\scr B}
\hbox{\rm $r$-body\ operator} \rket{\scr B},
\eeq
where the $r$-body operator in the spin--flavor Fock space is constructed
out of the fields and spurions in the effective lagrangian.
This form ensures that the $N$-counting of the effective lagrangian is
the same as that found in the previous section.

We now consider the question of the consistency of this effective
lagrangian.
Suppose we form a graph from two terms in the effective lagrangian,
corresponding to an $r_1$- and an $r_2$-body operator.
The graph will therefore have the form $1/N^{r_1 + r_2 - 2}$ times
an $(r_1 + r_2)$-body operator, apparently violating the $N$-counting of
the previous section.
Such contributions can only be consistent if the $(r_1 + r_2)$-body operator
has a special form, so that it can be written as a fewer-body operator.
We will show that this is in fact what happens in some examples, but we
have not been able to prove the consistency of this effective lagrangian in
general.

We first consider baryon-pion scattering at low energies.
For simplicity, we work in the chiral limit.
The three graphs that contribute are shown in fig.\ (3).
The graph in fig.\ (3c) gives a contribution of order $1/N$ to the
amplitude,\footnote{$^\dagger$}
{We work with non-relativistically normalized states, so that there is no
dependence on the mass in the normalization of amplitudes.}
while the graphs in fig.\ (3a,b) lead to an amplitude \GS\DM
\eq
\label\twopiscat
\scr A(\scr B \pi_1^{A} \to \scr B' \pi_2^B) \propto
\frac{q_1^j q_2^k}{q^0 f_\pi^2}
\rbra{\scr B'} [ X^{Aj}, X^{Bk} ] \rket{\scr B} + \cdots,
\eeq
where
\eq
X^{jA} \equiv (\lam^A)^a_b (J^j)^\al_\beta \al^\dagger_{a\al} \al^{b\beta},
\eeq
is the axial current in the spin--flavor Fock space corresponding to
the pion $\pi^A$.
Because the initial and final baryons are on shell, their residual momenta
are $O(1/N)$, and can be neglected.
Since $\rbra{\scr B'} X^{Aj} \rket{\scr B} \sim N$ and
$f_\pi \sim \sqrt N$ \Witten, one na\"\i vely expects $\scr A \sim N$,
which would violate unitarity for any fixed kinematics as $N\to\infty$.
However the commutator of two axial currents is a vector current, so for
baryons with $J\sim 1$ the term shown in eq.\ \twopiscat\ is of order
$1/N$.
If we include the mass splitting between baryons with different spins,
we obtain a contribution to the amplitude proportional to
\eq
\frac 1{f_\pi^2}\, \frac 1N\, \rbra{\scr B'} X^{Aj} \op{J^2} X^{Bk}
+ X^{Bk} \op{J^2} X^{Aj} \rket{\scr B} \sim 1,
\eeq
and so the full amplitude is actually of order 1 in the $1/N$
expansion.\footnote{$^\dagger$}
{We thank A. V. Manohar for pointing this out to us.}

Similarly, we can consider scattering processes involving more pions.
The graphs shown in fig.\ (4) give rise to a scattering amplitude
\eq
\label\threepi
\eqalign{
\scr A(\scr B \pi_1^A \to \scr B' \pi_2^B \pi_3^C) \propto
\frac{q_1^j q_2^k q_3^\ell}{q_1^0 q_2^0 q_3^0 f_\pi^3}
\biggl(& q_2^0
\rbra{\scr B'}[X^{Aj},[X^{Bk}, X^{C\ell}]] \rket{\scr B} \cr
&\quad - q_1^0 \rbra{\scr B'}[X^{Bk},[X^{C\ell}, X^{Aj}]]\rket{\scr B}
\biggr) + \cdots. \cr}
\eeq
The na\"\i ve estimate of this term is $\scr A \sim N^{3/2}$, but the
double commutators lead to matrix elements of the form
$\rbra{\scr B'} X \rket{\scr B} \sim N$, so that in fact
$\scr A \sim 1 / \sqrt N$ \DM.
The finiteness of this amplitude in the large-$N$ limit was used in
ref.\ \DM\ to argue that there are no $1/N$ corrections to the axial
current matrix element.
It is easy to check that the corrections to the axial current that we
have enumerated give rise to corrections to this amplitude which are
consistent in the large-$N$ limit.

In both of the cases considered above, products of operators that
na\"\i vely violate the $N$-counting rules had a commutator structure that
allowed them to be expressed in terms of operators with fewer creation
and annihilation operators.
As argued above, a mechanism of this sort is required for consistency
in the large-$N$ limit.
This commutator structure has its origin in the antisymmetry of tree
amplitudes under crossing.
For loop amplitudes, the ``$i\ep$'' part of the baryon propagator
ruins this antisymmetry, and amplitudes are no longer proportional
to commutators.
For example, if the pions have a common mass $m_\pi$, the baryon masses
have a 1-loop contribution \Jenkins
\eq
\label\mshift
\Delta M_B \propto \frac{m_\pi^3}{16\pi f_\pi^2}\,
\rbra{\scr B} X^{Aj} X^{Aj} \rket{\scr B}
\propto \frac{m_\pi^3}{16\pi f_\pi^2}\, \left[ N^2 + J(J+1) \right]
\eeq
for a baryon with spin $J$ (fig.\ (5)).
The leading term grows with $N$, but it is a harmless constant mass
shift consistent with the $N$-counting results given in table 1.
The mass difference between baryons with different spin is of the form
derived in section 1.

We have verified that all of the examples considered in
refs.\ \GS\DM\Jenkins\ give consistent results using our approach.
However, other than the remarks made above, we do not have any general
insight into the cancellation mechanism at work.
However, we believe that the effective lagrangian we have written down is
the most general one compatible with the symmetries and $N$-counting
structure of the large-$N$ limit.
We believe that {\it any} effective lagrangian which is defined by its field
content and symmetry structure is consistent as long as all allowed terms
are included.
(This ``theorem'' is advocated by Weinberg as the foundation of the
method of effective lagrangians
\ref\Weinberg{S. Weinberg, \PA{96}{327}{1979}.}.)
If this is correct, the effective lagrangian we have described provides
an explicit and consistent description of pion--baryon interactions at
low energies in the $1/N$ expansion.

\section{The Non-relativistic Constituent Quark Model}

Traditionally, $SU(6)$ relations have been justified by appealing to a
non-relativistic constituent quark model (NRCQM) in which the $SU(6)$
symmetry arises because the constituent quarks are heavy.
In this section, we make some brief remarks comparing our results to
what is expected from such a model.
We include this discussion to point out
that the pattern of corrections to $SU(6)$ relations we have derived above
is very different from what is expected from a NRCQM.

The $N$-counting of the NRCQM is identical to that of QCD, and so
$SU(6)$-violating amplitudes will be suppressed by powers of
$1/N$ in the NRCQM.
However, when we discuss the NRCQM below, we will {\it not} treat $1/N$ as
a small parameter, since the point of our discussion is whether the $SU(6)$
symmetry can be explained {\it solely} in terms of the heaviness of the
constituent quarks.

The NRCQM can be formulated by assuming that below the
QCD chiral symmetry breaking scale but above the confinement scale, QCD
dynamics can be approximated by an effective lagrangian containing
constituent quarks and weakly-interacting gluons
\ref\ChQM{For a modern exposition, see H. Georgi and A. V. Manohar,
\NPB{234}{189}{1984}.}.
The constituent quarks are to be viewed as composite objects whose
properties are the result of non-perturbative QCD dynamics at the chiral
symmetry breaking scale.
The most important difference between the constituent quarks and the
``current'' quarks that appear in the fundamental QCD lagrangian is that
the constituent quarks have a flavor-independent mass
$M_Q \sim \Lam_{\rm QCD}$ due to chiral symmetry breaking.
Because chiral symmetry is broken spontaneously, the theory also contains
$8$ ``pions'' in the chiral limit.

In this model, it is easy to see how the $SU(6)$ spin--flavor
symmetry arises.
If we assume that the effective gluon coupling $\al_s$ is weak, then the
constituent quarks inside a baryon will be nonrelativistic with binding
energy $E_B \sim \al_s^2 M_Q \ll M_Q$.
The interactions of the constituent quarks at energies of order $E_B$
therefore exhibit an $SU(6)$ spin--flavor symmetry exactly analogous to the
spin--flavor symmetry discussed recently for $b$ and $c$ quarks
\ref\heavy{N. Isgur and M. Wise, \PLB{232}{113}{1989};
{\bf 237B}, 527 (1990); H. Georgi, \PLB{240}{447}{1990}.}.
This $SU(6)$ symmetry will give rise to the same relations for baryon
masses and matrix elements derived above in the large-$N$ limit of QCD.

However, it should be emphasized that the picture of QCD dynamics contained
in the NRCQM  has some serious drawbacks when taken seriously.
First of all, if the constituent quarks are heavy and weakly interacting,
there should be weakly bound states with the same quantum numbers as the
pions with masses $\simeq \frac 23$ of a baryon mass.
Furthermore, because of the $SU(6)$ symmetry, these states should lie
in an $SU(6)$ multiplet with a nonet of vector mesons.
Identifying these states with observed mesons is rather problematic.
Another difficulty with the constituent quark model is that it leads us
to expect hadron form factors to deviate significantly from point-like
behavior for momentum transfers of order $E_B \ll M_Q \simeq 300 \MeV$.
Instead, we find that baryon form factors vary on a scale of order
$m_\rho \simeq 770 \MeV$.
Finally, the gluons which couple the constituent quarks confine at
some scale below $E_B$ (since the gluons are supposed to be weakly
interacting at that scale), so we expect glueball states with mass
$\ll M_Q$.
It is not clear whether any of these objections is fatal, but we want
to emphasize that the validity of the NRCQM should not be taken for granted,
despite its successes.

Leaving these problems aside, we want to compare the expected pattern of
corrections to the $SU(6)$ relations from the NRCQM to what we found for
the $1/N$ corrections in QCD.
This is easy to do if we write an effective lagrangian for the constituent
quarks treated as heavy fields:
\eq
\scr L_{\rm NRCQM} = \mybar Q iv\cdot D Q +
g_Q \mybar Q A\cdot s Q +
\frac b {2M_Q} \mybar Q v^\mu s^\nu \twi G_{\mu\nu} Q +
O(1 / M_Q^2).
\eeq
Here $Q$ is the constituent quark field, $D_\mu$ is the gluon covariant
derivative, and $\twi G_{\mu\nu} \equiv \ep_{\mu\nu\lam\rho} G^{\lam\rho}$
is the dual of the gluon field strength.
The $SU(6)$ symmetry is violated by the pion couplings and by the color
magnetic moment term.
Pion loops give rise to $SU(6)$-breaking corrections to matrix elements
which vanish at zero momentum in the chiral limit.
When the current quark masses are turned on, the pion loops give
$SU(6)$-breaking  corrections at zero momentum proportional to powers of the
current quark masses.
The $1/M_Q$ terms give rise to $SU(6)$-breaking corrections at zero
momentum.
These corrections will be proportional to powers of the $SU(6)$-breaking
spurion $s^\mu / M_Q$;
only even powers of this spurion can appear in corrections by parity
invariance.
Thus, {\it all} of the $SU(6)$ relations for matrix elements and baryon
masses are expected to have $1/M_Q^2$ corrections in the NRCQM.
This pattern of corrections is very different from that found in the
$1/N$ expansion of QCD, where quark-model relations can have corrections
which are either $1/N$ or $1/N^2$.

\section{Conclusions}

In this paper we have developed a diagrammatic expansion for
baryons in the large-$N$ limit.
This method gives a simple recipe for determining the $N$-dependence
of physical quantities.
Using these techniques, we have shown that forward matrix elements of
quark bilinears obey $SU(2N_F)$ spin--flavor relations.
Our approach makes a direct connection between the emergence of these
relations and the spin--flavor properties of the baryon wavefunctions,
which we feel is very attractive.

On a more practical level, our methods allow the enumeration of the
$1/N$ corrections to the large-$N$ results to all orders in $1/N$,
completing previous partial results on the classification of $1/N$
corrections.
We have also written down explicitly an effective lagrangian which
describes the interactions of baryons and pions at low energies in the
$1/N$ expansion.
The consistency of this lagrangian in the large-$N$ limit is not manifest,
because some of the couplings grow with $N$.
While we have not been able to prove consistency in general, we have
verified it in a number of cases and given general arguments indicating
that this effective lagrangian must in fact be correct.

After completing this paper, we received
ref.\ \ref\Havad{C. Carone, H. Georgi, and S. Osofsky, Harvard preprint
HUTP-93/A032},
which considers baryon masses and matrix elements using the Hartree--Fock
approach of ref.\ \Witten.
We also received
ref.\ \ref\UCSD{R. Dashen, E. Jenkins, and A. V. Manohar, UCSD preprint
UCSD/PTH 93-21, hep-ph 9310379.}\ which develops further the approach of
refs.\ \GS\DM.

\section{Acknowledgements}

We would like to thank J. Anderson, K. Bardakci, H. Murayama, R. Sundrum,
and M. Suzuki for helpful discussions while this work was in progress.
We are also very grateful to A. V. Manohar for pointing out an error in an
earlier version of this paper, and A. Cohen for discussions on related
matters.
JMR would like to thank the U.S. Department of Energy for a
Distinguished Postdoctoral Research Fellowship.
This work was supported by the Director, Office of Energy
Research, Office of High Energy and Nuclear Physics, Division of High
Energy Physics of the U.S. Department of Energy under Contract
DE-AC03-76SF00098.

\vfill\eject
\appendix{A}{Classification of Operators}

In this appendix, we consider the classification of few-body operators with
given transformation properties under \sf.
Since only the spin and flavor quantum numbers are relevant for our
discussion, it is useful to use a notation in which only spin and flavor
indices appear.
To do this, we note that the fact that the states $\bk$ are antisymmetrized
in color allows us to treat the quarks as colorless bosons.
To make this precise, we make use of the identity
\def\under#1#2{\mathop{\phantom\dagger#1}
\limits_{\lower.3em\hbox{$\scriptstyle #2$}}}
\eq
\label\QCDalg
a^\dagger_{c\gam A} a^{b \beta A} \bk =
\scr B^{a_1 \al_1 \cdots a_N \al_N} \ep^{A_1 \cdots A_N}
\sum_{s = 1}^N \del^b_{a_s} \del^\beta_{\al_s}
\under{a^\dagger_{a_1 \al_1 A_1}}{1} \cdots
\under{a^\dagger_{c \gam A_s}}{s} \cdots
\under{a^\dagger_{a_N \al_N A_N}}{N} \ket 0.
\eeq
Exactly the same algebraic relation is satisfied by {\it bosonic} creation
and annihilation operators carrying no color.
Specifically, we define
\eq
\rket{\scr B} \equiv \scr B^{a_1 \al_1 \cdots a_N \al_N}
\al^\dagger_{a_1 \al_1} \cdots \al^\dagger_{a_N \al_N} \rket 0,
\eeq
where the $\al^\dagger$'s are bosonic creation operators which act on a
spin--flavor Fock space.
We use the ``curved bra--ket'' notation to distinguish these states from
QCD states. We then have
\eq
\label\QMalg
\al^\dagger_{c\gam} \al^{b\beta} \rket{\scr B} =
\scr B^{a_1 \al_1 \cdots a_N \al_N}
\sum_{s = 1}^N \del^b_{a_s} \del^\beta_{\al_s}
\under{\al^\dagger_{a_1 \al_1}}{1} \cdots \under{\al^\dagger_{c\gam}}{s} \cdots
\under{\al^\dagger_{a_N \al_N}}{N} \rket 0.
\eeq
Comparison of eqs.\ \QCDalg\ and \QMalg\ shows that
\eq
\label\Qcorr
\bb \hat{\scr O}^{(r)} \bk \propto
\rbra{\scr B} \scr O^{(r)} \rket{\scr B},
\eeq
where we use the caret notation to denote QCD operators, and
\eq
\label\QMop
\scr O^{(r)} \equiv
X^{a_1 \al_1 \cdots a_r \al_r}_{b_1 \beta_1 \cdots b_r \beta_r}
\al^\dagger_{a_1 \al_1} \al^{b_1 \beta_1} \cdots
\al^\dagger_{a_r \al_r} \al^{b_r \beta_r}.
\eeq

Our job is now to classify the general $r$-body operators with
specified \sf\ transformation properties.
First note that
the most general $r$-body operator can be written as
\eq
\label\classop
\scr O^{(r)} =
X^{a_1 \al_1 \cdots a_r \al_r}_{b_1 \beta_1 \cdots b_r \beta_r}
\al^\dagger_{a_1 \al_1} \cdots \al^\dagger_{a_r \al_r}
\al^{b_1 \beta_1} \cdots \al^{b_r \beta_r} + \cdots ,
\eeq
where the omitted terms arise from the re-ordering
of the creation and annihilation operators.
These omitted terms are irrelevant for the classification of the
operators, since they are $s$-body operators, with $s < r$.
We can therefore classify the operators inductively, starting with
1-body operators.
We will make frequent use of this simplification in this appendix.

With the ordering chosen above, the creation operators project out the
part of $X$ which is symmetric under simultaneous
interchange of spin and flavor indices
$a_1 \al_1 \leftrightarrow a_2 \al_2$, \etc.
For notational simplicity, we will consider the case of 3 flavors,
although the results are valid for an arbitrary number of flavors.

\subsection{Singlets}

In this subsection, we classify the $SU(2)_v \times SU(3)$ singlet
$r$-body operators.
We proceed inductively in $r$.
It is clear that the only 1-body singlet operator is
$\al^\dagger_{a\al} \al^{a\al}$.

Now suppose that we have classified all of the singlet operators up to
the $(r - 1)$-body operators, and consider the $r$-body operators.
The tensor $X$ in eq.\ \classop\ must be a linear combination of products
of the $SU(3)$ and $SU(2)_v$ invariant tensors
$\del^a_b$, $\ep^{abc}$, $\del^\al_\beta$, and $\ep^{\al\beta}$.
First we note that the spin and flavor $\ep$ symbols always appear in
pairs and can thus be replaced by Kronecker $\del$'s using eq.\ \epid.
Therefore, the tensor $X$ in eq.\ \classop\ can be written as a linear
combination of terms of the form
\eq
\label\delform
X^{a_1 \al_1 \cdots a_r \al_r}_{b_1 \beta_1 \cdots b_r \beta_r} =
\del^{a_1}_{b_1} \cdots \del^{a_r}_{b_r}
Y^{\al_1 \cdots \al_r}_{\beta_1 \cdots \beta_r},
\eeq
where we have used the symmetry properties of the creation and annihilation
operators to put the flavor indices in canonical order.
The tensor $Y$ is an $SU(2)_v$ singlet, and can therefore be written as a
linear combination of tensors of the form
\eq
\label\thew
Y^{\al_1 \cdots \al_r}_{\beta_1 \cdots \beta_r} =
\del^{\al_1}_{\beta_{\sig_1}} \!\cdots \del^{\al_r}_{\beta_{\sig_r}},
\eeq
where $\sig$ is a permutation of $1, \ldots, r$.
(We have used up our freedom to re-order the indices in writing
eq.\ \delform.)

We can rewrite $SU(2)_v$ tensors of the form eq.\ \thew\ by writing the
permutation $\sig$ as a product of simple interchanges.
In this way, $Y$ can be written as a product of the tensors
\eq
\label\thes
S^{\al_1 \al_2}_{\beta_1 \beta_2} \equiv \del^{\al_1}_{\beta_2}
\del^{\al_2}_{\beta_1}.
\eeq
We can then use the identity
\eq
\label\Jid
(J_j)^{\al_1}_{\beta_1} (J_j)^{\al_2}_{\beta_2} = \sfrac 12\!\bigl[
S^{\al_1 \al_2}_{\beta_1 \beta_2} - \sfrac 12 \del^{\al_1}_{\beta_1}
\del^{\al_2}_{\beta_2} \bigr]
\eeq
to write $Y$ as a linear combination of products of $SU(2)_v$ generators
and the identity matrix. Finally, some simple $SU(2)$ algebra gives
us an expression for
$\scr O^{(r)}$ as a linear combination of products of $\op{J_j}$'s and
$\op{1}$'s (the number operator), which can be reduced to the form
$\op{1}^{r - 2s} \op{J^2}^s$, where we have used the notation
\eq
\op{J_j} \equiv \al^\dagger_{a\al} (J_j)^\al_\beta \al^{a\beta},
\eeq
and $\op{J^2}\equiv \op{J_j} \op{J_j}$.

To understand these last steps in detail, consider a simple
example which can be easily extended to the general case:
\eq
Y^{\al_1 \al_2 \al_3}_{\beta_1 \beta_2 \beta_3} \equiv
\del^{\al_1}_{\beta_2} \del^{\al_2}_{\beta_3} \del^{\al_3}_{\beta_1}
= S^{\al_1 \gam_2}_{\beta_1 \beta_2} S^{\al_2 \al_3}_{\gam_2 \beta_3}.
\eeq
Using eq.\ \Jid, we can write the corresponding 3-body operator as
\eq
\label\othree
\scr O^{(3)} = 4 \op{J_j} \op{J_k J_j} \op{J_k}
- 2 \op{1}\op{J^2}
+ \sfrac 14 \op{1}^3 + \cdots.
\eeq
The ellipses in eq.\ \othree\ indicate 1- and 2-body operators obtained
by re-ordering the creation and annihilation operators.
Using the identity
\eq
\label\Jidtoo
J_j J_k = \sfrac 14 \del_{jk} + \sfrac i2 \ep_{jk\ell} J_\ell,
\eeq
we obtain
\eq
\scr O^{(3)} = -\op 1 \op{J^2}
+ 2i \ep_{jk\ell} \op{J_j} \op{J_k} \op{J_\ell}
+ \sfrac 14 \op{1}^3 + \cdots.
\eeq
Finally, we use the identity
\eq
\label\finid
\ep_{jk\ell} \op{J_k} \op{J_\ell} =
\sfrac 12 \ep_{jk\ell} \bigl[\op{J_k}, \op{J_\ell} \bigr] =
i \op{J_j}
\eeq
to write
\eq
\scr O^{(3)} = -3\op{1} \op{J^2} + \sfrac 14 \op{1}^3 + \cdots.
\eeq

These steps can be repeated for an arbitrary tensor $Y$ of the form of
eq.\ \thew.
After writing $Y$ as a product of the $S$ tensors defined in eq.\ \thes,
we use the identity eq.\ \Jid\ to write $\scr O^{(r)}$ as a
linear combination of products of $\op 1$ and operators of the form
$\op{J \cdots J}$, where the indices on the $J$'s are contracted using
$\del_{ij}$'s.
Using eq.\ \Jidtoo, we reduce this to products of $\op 1$'s
and $\op{J_j}$'s, contracted with $\del_{jk}$'s and $\ep_{jk\ell}$'s.
We then use eq.\ \epid\ to eliminate the $\ep_{jk\ell}$'s in pairs in favor
of $\del_{jk}$'s, and then apply eq.\ \finid\ to eliminate the last
$\ep_{jk\ell}$ (if any).
In this way, we obtain products of $\op{1}$'s and $\op{J_j}$'s contracted
with $\del_{jk}$'s, which can be written as $\op{1}^{r - 2s} (\op{J^2})^s$
up to operators with fewer creation and annihilation operators.
Thus, we have shown that the most general $r$-body operator is a linear
combination of operators of the form
\eq
\op{1}^{r - 2s} \op{J^2}^s.
\eeq

\vfill\eject
\subsection{Quark Bilinears}

In this subsection, we classify the possible operators with the quantum
numbers of the tensor $W^a_b \Gam^\al_\beta$, corresponding to the matrix
elements of a quark bilinear.
An arbitrary quark bilinear operator can be written as a linear combination
of operators of this form.
Our argument is very similar to the one presented above for singlets, and
so we will be somewhat terse.

We again proceed inductively.
We begin by considering the case $\tr W = \tr \Gam = 0$.
In this case, the only effective 2-body operator with the right quantum
numbers is
\eq
\scr O^{(2)} \equiv \al^\dagger_{a\al} W^a_b \Gam^\al_\beta \al^{b\beta}.
\eeq
Now assume we have classified all operators up to $(r - 1)$-body operators,
and consider the $r$-body operators.
A general $r$-body tensor of the form of eq.\ \classop\ with the
right quantum numbers can be written as a linear combination of
tensors of the form
\eq
\label\thefirstform
X^{a_1 \al_1 \cdots a_r \al_r}_{b_1 \beta_1 \cdots b_r \beta_r} =
W^{b'}_{a'} \Gam^{\beta'}_{\al'}
Y^{a' \al' a_1 \al_1 \cdots a_r \al_r}_{b' \beta' b_1 \beta_1
\cdots b_r \beta_r},
\eeq
where $Y$ is an $SU(2)_v \times SU(3)$ invariant tensor.
It must therefore consist of combinations of $\delta^a_b$'s,
$\delta^\al_\beta$'s, and $\ep$ symbols for the groups $SU(2)_v$ and
$SU(3)$.
However, since $Y$ has an equal number of upper and lower indices
for both groups, the $\ep$ tensors must appear in pairs, and we can use
eq.\ \epid\ to eliminate them in favor of combinations of $\del$'s.
The tensor in eq.\ \thefirstform\ thus reduces to a linear combination
of the form
\eq
\label\theform
X^{a_1 \al_1 \cdots a_r \al_r}_{b_1 \beta_1 \cdots b_r \beta_r} =
W^{a_1}_{b_1} \del^{a_2}_{b_2} \cdots \del^{a_r}_{b_r}
\Gam^{\al_s}_{\beta_t} \twi Y^{\al_1 \cdots \widehat{\al_s}
\cdots \al_r}_{\beta_1 \cdots \widehat{\beta_t} \cdots \beta_r},
\eeq
where $s,t = 1, \cdots, r$, and the caret denotes omission.
$\twi Y$ is an $SU(2)_v$ invariant tensor, formed from a linear
combination of products of $\del^\al_\beta$'s,
where the lower indices are permuted relative to their canonical order.
Going through the same steps as for the singlet operators, we can write
this tensor in terms of $\del^\al_\beta$'s and $J_j$'s
with indices in canonical order.
In this way, an arbitrary $r$-body operator can be written as a linear
combination of operators of the form
\eq
\scr O^{(r)} = \op{W J \cdots J \Gam J \cdots J}
\op{J \cdots J} \cdots \op{J \cdots J}
\op 1 \cdots \op 1 + \cdots,
\eeq
where the $J$'s are contracted using $\del_{jk}$'s,
$\op{W \Gam}\equiv \alpha^\dagger_{a\al} W^a_b \Gam^\al_\beta
\alpha^{b\beta}$,
and we have discarded all lower-body operators.

We now use the identity eq.\ \Jidtoo\ to reduce this to a product of
$\op 1$'s and $\op{J_j}$'s contracted with $\del_{jk}$'s and
$\ep_{jk\ell}$'s.
If $\Gam = J_j$, the operator is a linear combination of
\eqa
\label\firstop
\scr O^{(r)} &\sim \op{W J_j}
\op J \cdots \op J , \eol
&\qquad \op W \op{J_j}
\op J \cdots \op J , \eol
&\qquad \op{W J} \op{J_j}
\op J \cdots \op J , \eol
\label\secondlastop
&\qquad \ep_{j\cdots} \op W
\op J \cdots \op J , \eol
\label\lastop
&\qquad \ep_{j\cdots} \op{W J}
\op J \cdots \op J . \eeol
\eeq
The unlabeled $J$'s in eqs.\ \firstop--\lastop\ are contracted using
$\del_{jk}$ and $\ep_{jk\ell}$.
We have dropped all factors of the operator $\op 1$, since $\op 1 = N$ on
the states we are interested in, and these factors therefore do not affect
the $N$-counting (see eq.\ \opncount).

We now eliminate the $\ep$ symbols pairwise using the identity eq.\ \epid.
This leaves at most one $\ep$ symbol.
The identity eq.\ \finid\ can be used to eliminate the final $\ep$
symbol (if any) from eqs.\ \firstop--\secondlastop.
The only non-trivial way an $\ep$ can appear in eq.\ \lastop\ is
as a factor of
\eq
\label\badop
\ep_{jk\ell} \op{W J_k} \op{J_\ell} =
i\bigl[\op{W J_j}, \op{J^2} \bigr] + \cdots,
\eeq
where the ellipses denote 1-body operators.
\ignore
(Note that the \rhs\ appears to be a 3-body operator, but one can easily
see that it is a 2-body operator be commuting all of the creation operators
to the left.)
\endignore
However, the \rhs\ of eq.\ \badop\ can be seen to have the opposite
transformation property as $\op{W \Gam}$ under time reversal, so
this operator cannot appear.

In this way, we find that the most general $r$-body operator with the
quantum numbers of $W \Gam$ is a linear combination of the
operators
\eq
\label\bicorr
\op{W \Gam}, \quad
\op W \op{\Gam}, \quad
\op{W J_j} \op{J_j} \op \Gam,
\eeq
multiplied by an $SU(2)_v \times SU(3)$ singlet operator.
(We may need to add or subtract the hermitian conjugate from the operators
in eq.\ \bicorr\ so that the effective operator has the same hermiticity
properties as the corresponding QCD operator.)

It remains only to consider the case where $W = 1$ and/or $\Gam = 1$.
Repeating the above argument with trivial modifications shows that
eq.\ \bicorr\ gives the most general corrections in this case as well.


\listrefs

\vfill
\eject

\centerline{\bf Figure Captions}
\vskip .4in
\noindent
Fig.\ 1. (a) The Young tableaux describing the \sf\ representation
of the baryon states with spin $J$.
(b) The spectrum of baryon states in the large-$N$ limit.
The mass splitting $\Delta M$ between adjacent states with spin $J\sim 1$
is of order $1/N$, while for $J\sim \sqrt{N}$, $\Delta M\sim 1$;
our methods do not allow us to study baryon states with $J \sim N$.
\smallskip
\noindent
Fig.\ 2. Typical diagrams contributing to the diagrammatic expansion of
$Z$ in eq.\ \zdef.
(a) A ``vacuum'' graph arising from the fully contracted
part of the Wick expansion. (b) A 3-body operator arising from the
normal-ordered terms in the Wick expansion.
The external creation and annihilation operators are denoted
by dashed lines.
The solid internal lines denote the usual quark propagators.
(c) A disconnected diagram with external annihilation
and creation operators. The entire contribution is normal-ordered.
(d) A typical diagram involving the ``hole'' propagator $S_H(x-y)$
(denoted by an internal dashed line).
\smallskip
\noindent
Fig.\ 3. The leading contributions to the amplitude for the process
$B\pi \rightarrow B'\pi$.
The diagrams (a) and (b) are proportional to axial current matrix elements
(see eq.\ \twopiscat).
Diagram (c) is proportional to the matrix element of a vector current, and
is of order $1/N$ for baryons with $J \sim 1$.
\smallskip
\noindent
Fig.\ 4. Contributions to the amplitude for
$B\pi \rightarrow B'\pi\pi$ which are proportional to matrix elements of
three axial currents.
\smallskip
\noindent
Fig.\ 5. A 1-loop diagram contributing to the baryon masses.

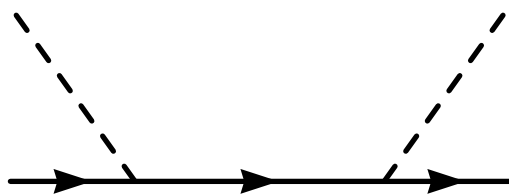
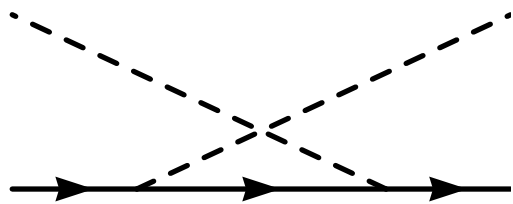
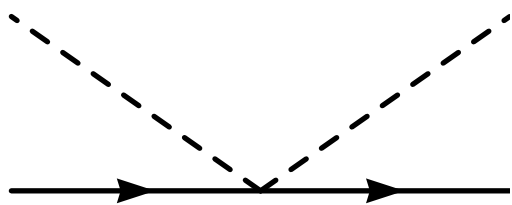
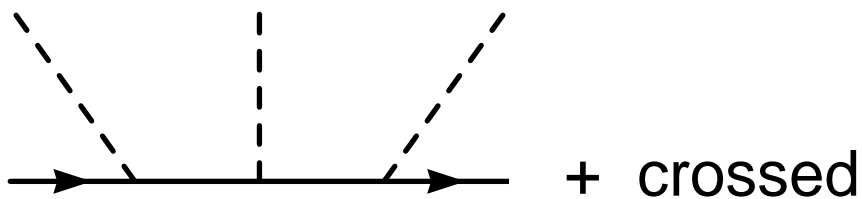
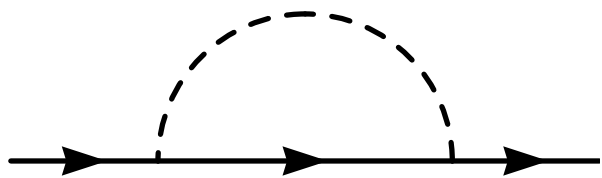
\bye